\documentclass{vgtc}

\graphicspath{{./}}

\usepackage{times}

\usepackage[utf8]{inputenc}

\usepackage{mathptmx}

\usepackage{graphicx}
\usepackage[inkscape=false,inkscapepath=basedir]{svg}
\usepackage{tabularx}
\usepackage[svgnames]{xcolor}
\usepackage{caption}
\usepackage{subcaption}
\usepackage{microtype}


\newcommand{\Iconemp}  {\includesvg[height=.9em]{phd-hat.svg}}
\newcommand{\Iconlead} {\includesvg[height=1em]{crown.svg}}
\newcommand{\Iconart}  {\includesvg[height=.8em]{artist-hat.svg}}
\newcommand{\Iconunk}  {\includesvg[height=1em]{magician-hat.svg}}

\newcommand{\WIconemp}  {\includegraphics[height=.9em]{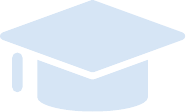}}
\newcommand{\WIconlead} {\includegraphics[height=1em]{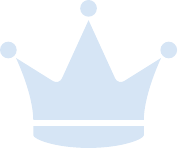}}
\newcommand{\WIconunk}  {\includegraphics[height=1em]{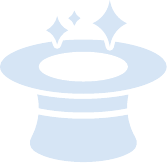}}

\newcommand{\LBIconemp}  {\includegraphics[height=.9em]{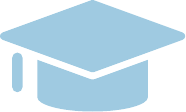}}
\newcommand{\LBIconlead} {\includegraphics[height=1em]{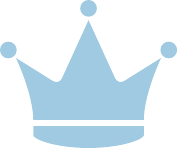}}
\newcommand{\LBIconart}  {\includegraphics[height=.8em]{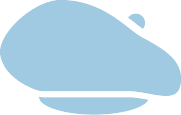}}
\newcommand{\LBIconunk}  {\includegraphics[height=1em]{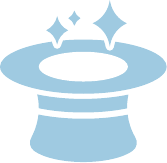}}

\newcommand{\BIconemp}  {\includegraphics[height=.9em]{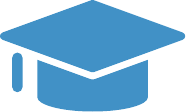}}
\newcommand{\BIconlead} {\includegraphics[height=1em]{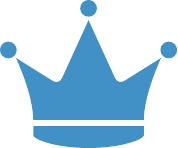}}
\newcommand{\BIconart}  {\includegraphics[height=.8em]{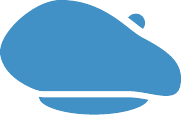}}
\newcommand{\BIconunk}  {\includegraphics[height=1em]{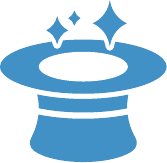}}

\newcommand{\DBIconemp}  {\includegraphics[height=.9em]{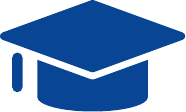}}
\newcommand{\DBIconlead} {\includegraphics[height=1em]{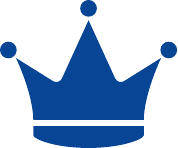}}
\newcommand{\DBIconart}  {\includegraphics[height=.8em]{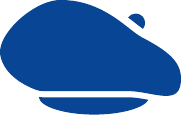}}
\newcommand{\DBIconunk}  {\includegraphics[height=1em]{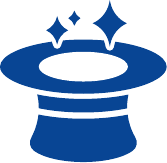}}

\definecolor{bulletone}{HTML}{084594}
\definecolor{bullettwo}{HTML}{4191c6}
\definecolor{bulletthree}{HTML}{9fcae1}
\definecolor{bulletfour}{HTML}{d6e5f5}
\newcommand{\opbulletone}{\raisebox{-.1ex}{\scalebox{1.5}{\textcolor{bulletone}{$\bullet$}}}}
\newcommand{\opbullettwo}{\raisebox{-.1ex}{\scalebox{1.5}{\textcolor{bullettwo}{$\bullet$}}}}
\newcommand{\opbulletthree}{\raisebox{-.1ex}{\scalebox{1.5}{\textcolor{bulletthree}{$\bullet$}}}}
\newcommand{\opbulletfour}{\raisebox{-.1ex}{\scalebox{1.5}{\textcolor{bulletfour}{$\bullet$}}}}

\newcommand{\formatTerm}[1]{\textsc{#1}}

\onlineid{0}

\vgtccategory{Research}

\vgtcinsertpkg

\crefname{figure}{Fig.}{Figs.}
\Crefname{figure}{Fig.}{Figs.}
\crefname{subfigure}{Fig.}{Figs.}
\Crefname{subfigure}{Fig.}{Figs.}
\crefname{table}{Table}{Tables}
\Crefname{table}{Table}{Tables}
\crefname{section}{Sec.}{Secs.}
\Crefname{section}{Sec.}{Secs.}
\crefname{subsection}{Sec.}{Secs.}
\Crefname{subsection}{Sec.}{Secs.}
\crefname{subsubsection}{Sec.}{Secs.}
\Crefname{subsubsection}{Sec.}{Secs.}

\title{Guidelines Are Not Rules: Characterizing Terminologies around Visualization Design Guidelines}

\author{Anna L. Chinni\thanks{e-mail: alchinni@wisc.edu}\\
     \scriptsize University of Wisconsin-Madison
\and Md Dilshadur Rahman\thanks{e-mail: dilshadur.rahman@utah.edu}\\
     \scriptsize University of Utah
\and Bon Adriel Aseniero\thanks{e-mail: bon.aseniero@autodesk.com}\\
     \scriptsize Autodesk Research
\and Petra Isenberg\thanks{e-mail: petra.isenberg@inria.fr}\\
     \scriptsize Univ.-Paris-Saclay, CNRS, Inria, LISN%
\and Kushin Mukherjee\thanks{e-mail: kushinm@cs.stanford.edu}\\
        \scriptsize Stanford University
\and Ghulam Jilani Quadri\thanks{e-mail: quadri@ou.edu}\\
        \scriptsize University of Oklahoma
\and Paul Rosen\thanks{e-mail: paul.rosen@utah.edu}\\
        \scriptsize University of Utah
\and Karen B. Schloss\thanks{e-mail: kschloss@wisc.edu}\\
        \scriptsize University of Wisconsin-Madison
\and Daniel Weiskopf\thanks{e-mail: weiskopf@visus.uni-stuttgart.de}\\
        \scriptsize University of Stuttgart
}

\teaser{
  \centering
  \includegraphics[width=0.9\linewidth]{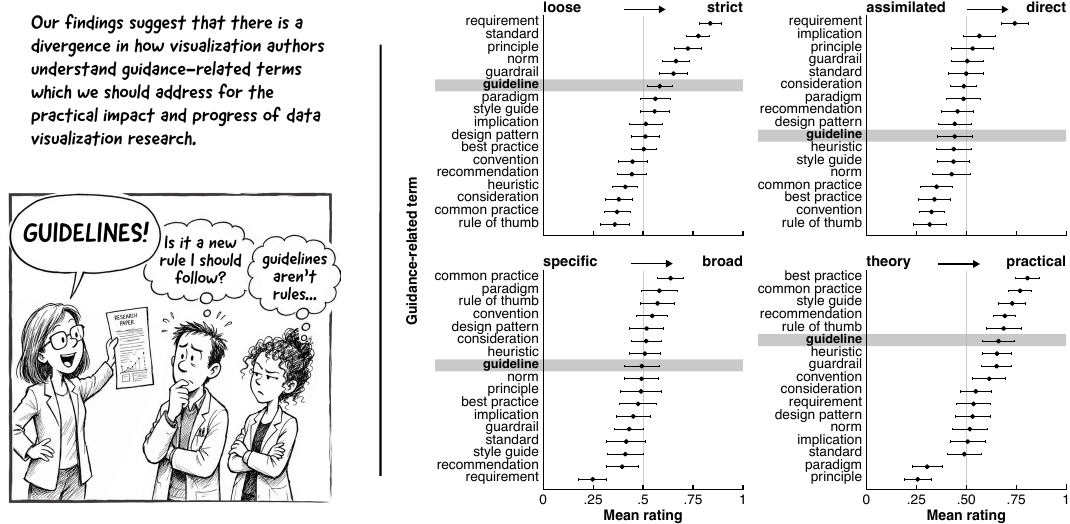}
  \caption{In visualization research, the term \formatTerm{guideline} is widely used to describe findings, yet the term is ambiguous and there is opportunity for more nuanced terminology. The ratings on the right show results from our crowdsourced study of guidance-related terms (see \Cref{sec:study2}) (ordered by mean ratings averaged across participants with 95\% confidence intervals) across 4 dimensions.
  }
  \label{fig:teaser}
}

\abstract{
A common expectation is that visualization research will produce recommendations for how researchers and practitioners take action or make design decisions.
We often express these as ``guidelines.''
Yet, the term ``guideline'' is both ambiguous and loosely defined, and what one researcher considers a guideline may be too broad, too loose, or too strict for another. We take a closer look at a broader set of terms that can express results around visualization research and untangle how these words are understood in the community in relation to other similar terms. We base our work on an exploratory study with experts, followed by a crowdsourcing study with a separate mapping phase ($n=30$) and rating phase ($n=42$) targeting input from the broader visualization community, and an analysis of the use of terminology in 3,877 IEEE VIS papers published from 1990 to 2024.
Based on our findings, we call for more nuanced, precise discussions of research outcomes and their communication to the broader community, including practitioners and students.

}


\begin{document}

\firstsection{Introduction}

\maketitle

In the field of visualization, there exists a desire and need for generalizable visualization research findings  that directly inform practice. Prescriptive advice on ``what to do when'' finds its way into introductory courses, textbooks, or tutorials, and is specifically common in practitioner-oriented visualization literature. It is, therefore, not surprising that an (erroneous) expectation is growing in the community that all forms of research should produce generalizable results that become quickly applicable and adoptable by the community. We have seen the expectation increasingly raised by reviewers who ask for guidelines for design---even when guidelines were inappropriate to propose from the current state of the work.

One reason for this mismatch in expectations may be a misunderstanding or disagreement about the meaning of guidance-related terminology in the community. Some may, for example, consider a \formatTerm{guideline} as a broad and loose \formatTerm{rule of thumb}, while others see it as a strict rule to be followed \cite{chen:2017:PathwaysTheoreticalAdvances,sedig2016DesignVisualizationsHumanInformation}. The importance of shared, consistently used terminology is often discussed in the study of science because it is crucial to how a community accumulates knowledge and community members learn to conduct science themselves \cite{Kuhn:1970:StructureScientificRevolutions}. Shared terminology allows work to be comparable; if one set of papers derives guidelines from authors' opinions and another from rigorously validated empirical research, the resulting sets of guidelines become almost impossible to compare.

Ideally, authors would carefully choose or at least clearly define how they use the terms to describe the applicability of their research results, but we argue that this is rarely the case for the broader class of terms surrounding guidelines. Shared definitions have potential to reduce conflicts between authors and reviewers and help clarify expectations. Shared definitions would also further support cumulative knowledge building, literature searches, and meta-analyses of contributions. In discussions that began at a Dagstuhl Seminar~\cite{meyer2026navigating}, we set out to uncover the extent of the mismatch between community members' understanding and use of guidance-related terms. We engaged in in-depth discussions among visualization experts on a set of identified words and their meanings. We also reviewed the literature on the use of the identified terms and conducted two community surveys to assess how other visualization experts understood them.

Through studies with crowdsourced participants and of visualization literature, we found that the terms vary in how they are understood, and that this variability carries over into practice. In our crowdsourced studies, guidance-related terms separated along several dimensions, including strictness, breadth, evidentiary basis, and theoretical vs. practical orientation. Yet \formatTerm{guideline}, the term most often invoked in this debate, consistently fell near the center of these dimensions (\Cref{fig:teaser}),
suggesting that the field's most common term for guidance may also be its least settled in meaning. A complementary analysis of three decades of IEEE VIS publications showed a parallel pattern: authors rely on a narrow handful of terms, chiefly \formatTerm{guideline}, \formatTerm{recommendation}, and \formatTerm{consideration}, to describe outcomes from their work.
This narrow, ambiguous vocabulary has grown more common over the past decade rather than converging on more precise usage, underscoring that the mismatch between the nuance the community perceives and the words it actually uses is unlikely to resolve on its own.

Our findings suggest that, as we discuss ways to approach studying and producing effective visualizations, we should be mindful of the implications of the terms we use.
We hope our results may serve as a starting point for more nuanced, precise discussions of research outcomes and their communication to the broader community, including practitioners and students, while also supporting the systematic and functional articulation of guidelines that inform system development, guide research, and improve teaching visualization. The landscape of guidance terminology that we propose, therefore, provides a practical scaffolding for communicating research results and their relative utility.

\section{Related Work}
\label{sec:relatedwork}
As a discipline at the intersection of computer science, art, design, mathematics,
engineering, psychology, and many application domains, the visualization community has a very diverse relationship with ``rules''  or \formatTerm{guidelines}. When students of visualization classes are trained in computer science, they are often happy to learn established guidelines like Cleveland and McGill's ranking of elementary perceptual tasks \cite{cleveland1984GraphicalPerceptionTheory} but frustrated when they hear that these rankings do not universally apply \cite{davis:2024:RisksRankingRevisiting}. This frustration is perhaps unsurprising because one could argue that computer science training and practice are rule-based at their core; a focus on mathematics, algorithms, programming, or formal methods may instill a sense of certainty that rules are absolute, deterministic, and universally applicable. Art, and some areas of design, however, do not treat rules the same way. Here, rules are important for learning and mastering the craft but need to be broken to support creative expression \cite{packard:2023:ArtRulesHow}. Parsons and Shukla \cite{parsons:2022:ConsideringRoleGuidelines} argue that the visualization community is dominated by ``Technical Rationality,'' a shared view that practitioners' application of ``general, codified knowledge'' is important and researchers must deliver such knowledge. The authors further argue that the visualization field is predominantly positivist in nature, but that, to support design practice, both tacit and explicit knowledge must be better accounted for in the way visualization researchers communicate about and analyze their work.

Similar to Sedig and Parsons \cite{sedig2016DesignVisualizationsHumanInformation}, we argue that visualization as a discipline grapples with these differing views but needs to agree on precise, shared terminology to prevent conflicts among readers, authors, reviewers, and, eventually, practitioners who seek to learn from visualization research. In contrast to their work, we focus here on the guidance landscape: a set of terms used to describe the practical or theoretical outcomes of visualization research.

\subsection{Community Discourse on Guidelines}

A visualization community discourse on guidelines has been fostered by the VisGuides workshop series,
which started in 2016 under the name C4PGV \cite{Borgo:2019:C4PGVWebsite}.
The organizers identified a need to establish a centralized place to make guidelines and principles discoverable, describe their applicability, and record the evolution and improvement of principles and guidelines. By the fourth edition,
the term \formatTerm{principles} was dropped from the workshop title \cite{Bach:2022:VisGuides} and the workshop description focused on guidelines with little mention of principles. One of the main outcomes of the workshop series is the accompanying VisGuides website \href{https://visguides.org/}{https://visguides.org/} \cite{diehl2018visguides}.
Diehl et al.~\cite{diehl2020studying} later coded 248 posts that had been made on the site and found limited uptake and discussion, perhaps due to underlying difficulties with discussing details, searching for references, and abstracting and generalizing from specific empirical results to concrete advice. This issue recalls findings by Kim et al.~\cite{kim2026gap}, who documented a research--practice gap in visualization design guidelines through a mixed-methods study comparing 390 practitioner-facing guidelines with 235 empirical studies, and found frequent misalignment: empirical evidence often contradicted or only partially supported widely used guidelines, and researchers and practitioners prioritized different design attributes.

Another recent initiative, focused on guidelines, convened a Dagstuhl Seminar to unify visualization design recommendations~\cite{meyer2026navigating}.  The Dagstuhl Seminar was motivated by the question: ``How do we formulate and integrate the knowledge we produce to best serve the visualization community and the world broadly?'' and it included working groups on terminology (one of which is responsible for this paper), on practitioners, on competing value systems in visualization (this group held a panel at IEEE VIS 2025 \cite{Harrison:2025:VisPanel}), and on guidelines$+$AI.

Given efforts such as these and the many researchers who participated in workshops, seminars, or panels on guidelines, it is safe to say that the visualization community has increasingly examined the landscape of guidance. We argue, however, that a critical lens on the terminology remains missing. In fact, few papers have provided definitions of guidance-related terms. A notable exception is the work by Chen et al.~\cite{chen:2017:PathwaysTheoreticalAdvances}, who defined a guideline as ``a process or a set of actions that may lead to a desired outcome or, alternatively, actions to be avoided to prevent an undesired outcome''  and a principle as ``a law or rule that must be followed and is usually expressed in a qualitative description.'' These definitions broadly agree with ours, as offered in \cref{tab:terms-final}, but differ subtly in focus.

\subsection{The Origin of Guidelines}

Chen et al.~\cite{chen:2017:PathwaysTheoreticalAdvances} argued that guidelines for visualization stem from ``accumulated experience and knowledge about some causal relations in a process.'' If one views visualization research through a positivist lens, then this accumulated experience and knowledge must be grounded in objective empirical observation \cite{parsons:2022:ConsideringRoleGuidelines,saharan:2026:CriticalReflectionValues} and may lead to guidelines that ``can be systematized such that anyone can produce a high-quality, reliable visualization that shows the data without extraneous distractors'' \cite{saharan:2026:CriticalReflectionValues}. If one takes a more interpretivist viewpoint, then the experience and knowledge Chen et al.\ discuss are acquired through critical reflection on situated practice: observing how practitioners actually design \cite{saharan:2026:CriticalReflectionValues} or how people actually make sense of visualizations; and interpreting these observations in context. According to Meyer and Dykes~\cite{meyer:2019:CriteriaRigorVisualization}, visualization research has questions and goals that are suited to each of these viewpoints. Yet, it is important to understand and question the evidentiary basis of visualization guidance and the processes by which empirical findings become broadly accepted \cite{dourish2006implications,greenberg2008usability,kosara2016empire,lee2019broadening}.

These arguments have long emerged in human-computer interaction.
Greenberg and Buxton~\cite{greenberg2008usability}, for example, argued that empirical usability evaluations are often treated as a universal mechanism for validating research contributions, leading to design recommendations that are generalized beyond the scope of the supporting evidence. Similarly, Dourish~\cite{dourish2006implications} questioned the convention of translating empirical findings directly into design implications, arguing that empirical studies frequently contribute conceptual understanding rather than explicit design recommendations. According to Kosara~\cite{kosara2016empire}, many canonical visualization principles originate from empirical findings, expert opinion, and common practice that are generalized into broadly accepted recommendations without clearly communicating their evidentiary basis or intended scope. More broadly, Lee et al.~\cite{lee2019broadening} emphasized that visualization knowledge is generated through diverse research paradigms, suggesting that design guidance may arise from multiple forms of research contributions rather than a single empirical methodology.

Collectively, these works demonstrate that visualization guidance varies in its origin, evidentiary support, and intended scope. However, they do not distinguish among or define terms that may be most suitable for describing types of guidance in the visualization literature. This problem is acutely felt by researchers who attempt to structure and define the guideline space \cite{choi:2021:UnifiedFrameworkVisualization,diehl2018visguides,diehl2020studying}, use it in recommendation systems \cite{moritz:2019:FormalizingVisualizationDesign}, or derive theory from it.

\section{Studies on Guidance Terminology}

\label{sec:empirical}

We conducted two studies to deepen our understanding of the meaning of guidance terminology. The initial study took place during a Dagstuhl Seminar~\cite{meyer2026navigating}, in which we explored the space of guidance terminology and potential sources for such guidance (\Cref{sec:dagstuhl}). The second study built on this initial approach but used more formal embedding and dimensionality reduction procedures, using data collected from the broader visualization community (\Cref{sec:experiments}).

\subsection{Exploration of Guidance Terminology}
\label{sec:dagstuhl}

This section describes an informal but structured exploration of guidance-related terminology.
We aimed to build a shared foundation for understanding words used to describe different kinds of guidance and how those words vary in meaning among members of the visualization community. We also explored likely sources for providing different kinds of guidance.

\begin{table}[!b]
    \vspace{-3mm}
    \centering
    \caption{A list of definitions for the community to consider. The terms above the line were considered during the initial exploration (\Cref{sec:dagstuhl}). These reflect our definitions, modified by our study results and rewritten with the help of AI (\href{https://chatgpt.com/share/6a469678-9714-83eb-9762-08f90c0671a3}{https://chatgpt.com/share/6a469678-9714-83eb-9762-08f90c0671a3}).
    Icons indicate likely sources of each term, determined using survey data (see \Cref{sec:dagstuhl}):
    \includegraphics[height=.8em]{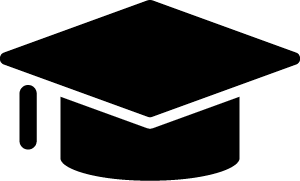}~empirical evidence;
    \includegraphics[height=.9em]{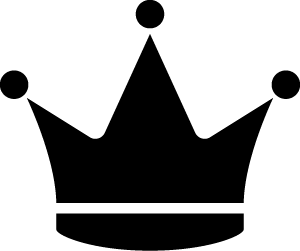}~eminent leader;
    \includegraphics[height=.7em]{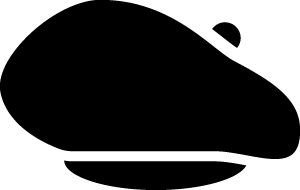}~art \& design community;
    \includegraphics[height=.8em]{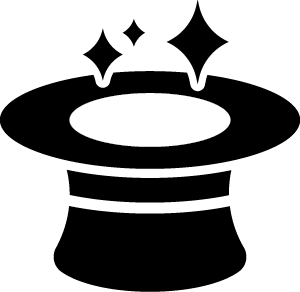}~emergent from unknown source (darker icons indicate more frequent responses: \opbulletfour~\opbulletthree~\opbullettwo~\opbulletone\,, see Supplemental \cref{fig:origin_mat} for data in matrix form).
    Terms below the line were added and used in Study~2 (\Cref{sec:experiments}) after the survey and, therefore, have no data on likely sources.
    }
    \label{tab:terms-final}

    \resizebox{\linewidth}{!}{
    \begin{tabularx}{1.33\columnwidth}{@{}>{\raggedleft\arraybackslash}p{7.5em}>{\raggedright\arraybackslash\hyphenpenalty=10000\exhyphenpenalty=10000}X@{\hspace{3pt}}c@{\hspace{3pt}}c@{\hspace{3pt}}c@{\hspace{3pt}}c@{}}
       \textbf{\formatTerm{Best Practice}}
         & A practical, experience-based method derived from assimilated knowledge and widely accepted for reliably producing desirable results.
         & {\BIconemp} & {\BIconlead} & {\DBIconart} & {\BIconunk}\\[2pt]
        \textbf{\formatTerm{Common Practice}}
         & A loose, broad, experience-based, and practical way of doing something, derived from assimilated knowledge and typically followed in practice.
         & {\WIconemp} & {\LBIconlead} & {\DBIconart} & {\DBIconunk}\\[2pt]
        \textbf{\formatTerm{Consideration}}
         & A loose factor or aspect to be kept in mind or given careful thought when thinking about or deciding something, without imposing any constraint or obligation.
         & {\DBIconemp} & {\DBIconlead} & {\DBIconart} & {\DBIconunk}\\[2pt]
        \textbf{\formatTerm{Convention}}
         & A rather loose, experience-based, and practical method of doing something, derived from assimilated knowledge and established in practice.
         & {\LBIconemp} & {\BIconlead} & {\DBIconart} & {\BIconunk}\\[2pt]
        \textbf{\formatTerm{Guideline}}
         & A moderately strict, practical recommendation derived from assimilated knowledge that helps steer actions or decisions without prescribing a rigid rule.
         & {\DBIconemp} & {\DBIconlead} & {\DBIconart} & {\BIconunk}\\[2pt]
        \textbf{\formatTerm{Guardrail}}
         & A strict, relatively specific, practical constraint intended to prevent undesirable results.
         & {\DBIconemp} & {\DBIconlead} & {\BIconart} & {\WIconunk}\\[2pt]
        \textbf{\formatTerm{Norm}}
         & A strict, widely accepted convention or practice grounded in assimilated knowledge and accumulated experience rather than formal theory.
         & {\BIconemp} & {\BIconlead} & {\DBIconart} & {\DBIconunk}\\[2pt]
        \textbf{\formatTerm{Paradigm}}
         & A theoretical, slightly broad framework or model that provides a modestly prescriptive way of approaching problems and developing solutions.
         & {\BIconemp} & {\LBIconlead} & {\LBIconart} & {\BIconunk}\\[2pt]
        \textbf{\formatTerm{Principle}}
         & A strict, theory-derived fundamental law or truth from which other concepts, rules, or conclusions are derived.
         & {\DBIconemp} & {\DBIconlead} & {\LBIconart} & {\LBIconunk}\\[2pt]
        \textbf{\formatTerm{Recommendation}}
         & A practical, experience-based, and rather loosely binding yet specific suggestion derived from assimilated knowledge about actions, choices, or considerations.
         & {\DBIconemp} & {\DBIconlead} & {\DBIconart} & {\WIconunk}\\[2pt]
        \textbf{\formatTerm{Rule of Thumb}}
         & A loose, broad, experience-based, and practical heuristic derived from assimilated knowledge and used for approximate decision-making.
         & {\BIconemp} & {\BIconlead} & {\DBIconart} & {\DBIconunk}\\[2pt]
        \textbf{\formatTerm{Standard}}
         & A strict, moderately specific set of established criteria intended to ensure desirable results.
         & {\LBIconemp} & {\WIconlead} & {\DBIconart} & {\LBIconunk}\\[2pt]
        \textbf{\formatTerm{Style Guide}}
         & A practical, slightly strict, and rather specific set of rules derived from assimilated knowledge and established usage to ensure consistency.
         & {\WIconemp} & {\DBIconlead} & {\DBIconart} & {\WIconunk}\\[2pt]
        \hline \\[-4pt]
        \textbf{\formatTerm{Design Pattern}} & A structured, reusable solution to a recurring design problem, grounded rather in assimilated knowledge and practice than in theoretical derivation. & & & & \\[2pt]
        \textbf{\formatTerm{Heuristic}} & A loose, experience-based, and practical approach derived from assimilated knowledge, used to guide discovery and problem-solving without strict formal rules. & & & & \\[2pt]
        \textbf{\formatTerm{Implication}} & The rather specific practical consequences, effects, or recommendations that follow from a piece of guidance or a set of findings. & & & & \\[2pt]
        \textbf{\formatTerm{Requirement}} & A strict and specific condition or obligation derived from empirical observation that must be satisfied.
         & & & & \\
    \end{tabularx}
    }
\end{table}

\subsubsection{Term Generation} \label{sec:term_generation}

To allow for nuanced and precise discussions of research outcomes, we need more than the term \formatTerm{guideline}. Six senior authors (BAA, PI, GJQ, PR, KBS, and DW) and one acknowledged contributor (CBX) discussed a purposefully large list of alternative terms to capture different aspects of the research results that might, for example, differ in their applicability, breadth, source, or strictness. We aimed to be forward-looking and include terms that are perhaps not yet used, but should be.
To produce the definitions in \cref{tab:terms-final} (also see the supplemental material for the preliminary definitions from the Dagstuhl discussion), we started with definitions from the Merriam-Webster English dictionary (online) and supplemented them with definitions from dictionary.com. We also asked ChatGPT for alternatives regarding specific nuances or variations of terms not captured by the dictionary definition and picked those that captured our intent.
We used AI when the dictionaries did not provide appropriate definitions; e.g.,
the two dictionaries defined \formatTerm{guardrail} only in its architectural sense. All authors reviewed and discussed the definitions and generated terms before inclusion in the final list.

\subsubsection{Exploring the Space of Terms} \label{sec:space_explore}
As we generated our list of terms, we began exploring their similarity space on a large post-it pad. We placed terms that seemed similar in meaning close together, and terms that seemed different far apart. We iterated through this process, moving the terms around, to try to understand not only which terms were similar/different, but to start to understand the underlying dimensions of this space. Early on, we established that one important dimension was likely \textit{strictness}, with terms like \formatTerm{guardrail} being near the \textit{strict} endpoint of the dimension and terms like \formatTerm{rule of thumb} being near the \textit{loose} endpoint of the dimension. We explored various possibilities of what a second dimension might be, and settled on \textit{coverage} ranging from \textit{broad} to \textit{specific}, with terms like \formatTerm{principle} being near the broad end and terms like \formatTerm{style guide} being near the specific endpoint.

\subsubsection{Studying the Dimensionality of Terms and Sources}
\label{sec:study1-dimensions}

Having established our list of terms and identifying dimensions in which these terms might vary%
, we next conducted an informal survey to better understand these dimensions. We invited the other participants of our seminar, not involved
in our working group; 13 responded. The survey had three parts. First, we presented participants with each of the 13 terms and asked them to rate each term on the following dimensions using a 5-point Likert scale: (1)~Strictness: the degree to which the term implies loose/suggestive vs.\ strict/enforced guidance, and  (2) Coverage: the degree to which the term spans the visualization design space, from very specific/narrow to very broad/general. Next, we gave the participants the list of terms again and asked them to categorize each term by its likely source, choosing from: empirical evidence (\raisebox{-0.2\height}{\Iconemp}), an eminent leader in the field (\raisebox{-0.2\height}{\Iconlead}), the art and design community (\raisebox{-0.2\height}{\Iconart}), or emergent from unknown sources (\raisebox{-0.2\height}{\Iconunk}). They were asked to select all sources that applied. Finally, we asked participants if there were any guideline-related terms that we left out that they thought we should consider.

\begin{figure}[t]
    \centering
    \includegraphics[width=.9\linewidth]{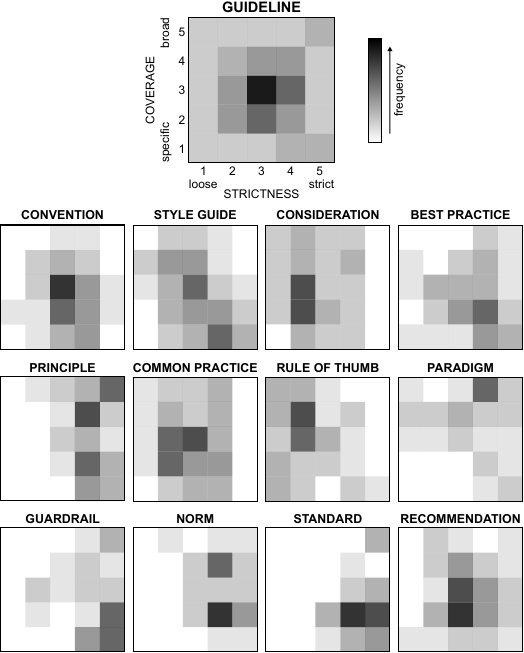}
    \caption{Frequency of term ratings along the dimensions of strictness (columns) and coverage (rows) from Study 1. Cell darkness indicates how many respondents chose that strictness\textendash coverage combination (darker indicates greater agreement). A tightly clustered dark region indicates strong consensus (e.g., \formatTerm{standard}); a diffuse, spread-out pattern indicates disagreement
    (e.g., \formatTerm{guideline}).
     }
    \label{fig:terms-grids}
    \vspace{-4mm}
\end{figure}

The heatmaps in \Cref{fig:terms-grids} show the frequency ratings along the dimensions of strictness and coverage (see caption for details).
Generally, judgments of the terms often covered the full spectrum, with \formatTerm{guidelines} being a term that most participants judged in the center of both scales (i.e., the black center) but also varied widely across the full range of the dimensions (i.e., ratings spread across the entire heatmap). Other terms, like \formatTerm{standard} and \formatTerm{guardrail}, however, were interpreted more consistently as rather specific and strict.

\begin{figure}[t]
    \centering
    \includegraphics[width=0.8\linewidth]{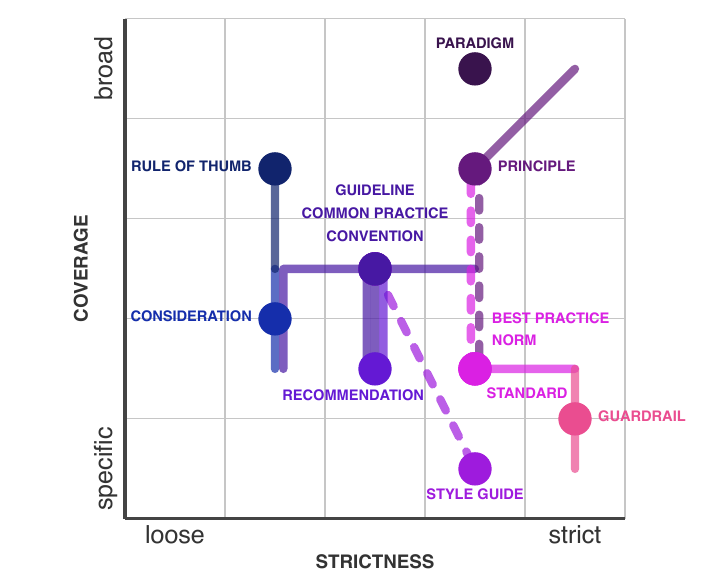}
    \vspace{-2ex}\caption{A visual map of how terms relate to one another, based on the data from the survey in \Cref{fig:terms-grids}, retaining the more frequent participant feedback (see text for details). Color coding: cooler colors = looser terms, warmer = stricter, lighter = more specific, \mbox{darker = more broad.}}
    \label{fig:term-uncertainty}
      \vspace{-4mm}
\end{figure}

To better understand how these terms relate to one another and are positioned in the Strictness/Coverage space, we superimposed the data from \Cref{fig:terms-grids} into a single plot in \Cref{fig:term-uncertainty}. We generated \Cref{fig:term-uncertainty} by (1)~only including cells from \Cref{fig:terms-grids} with greater than 25\% frequency for any term, (2)~marking the highest frequency cells for each term with a dot (or dots for ties), (3)~marking the remainder of cells which exceeded the 25\% frequency threshold with lines connecting to their corresponding dots (we used dashed lines when connected cells were not immediately adjacent), (4)~color-coding points and lines according to term strictness (cooler hues were looser, warmer hues were stricter) and broadness (lighter colors were more specific, darker colors were more broad).
The resulting \Cref{fig:term-uncertainty} serves as an initial map of how guidance-related terminology varies from term to term, and also how much individuals vary in their interpretation of each term. For example, \formatTerm{guideline} landed near the middle of both dimensions with considerable variability in strictness, which may suggest high ambiguity in how the community interprets this term. \formatTerm{Rule of thumb} was considered relatively loose and broad, though some participants thought it was more specific. In \Cref{fig:term-uncertainty}, Strictness increased from \formatTerm{consideration} to \formatTerm{recommendation} to \formatTerm{standard}, and all of these terms were moderate on coverage. And broadness increased from \formatTerm{guardrail} to \formatTerm{principle} to \formatTerm{paradigm}, but they were all relatively strict. These results suggest that when discussing guidance in the context of research, education, and design practice, there is opportunity to choose guidance terminology that aligns best with the intended strictness and coverage.

\Cref{tab:terms-final} shows the results regarding the likely source for each of the 13 terms. The source categorizations seemed to differ across terms, such as  \formatTerm{principle} stemming most empirical evidence (\raisebox{-0.2\height}{\Iconemp}) and eminent leaders (\raisebox{-0.2\height}{\Iconlead})  and \formatTerm{rule of thumb} emerging from the art and design community  (\raisebox{-0.2\height}{\Iconart}) or emerging from unknown sources (\raisebox{-0.2\height}{\Iconunk}).  These results may be useful for aligning terminology choices with the source of information.

Finally, in response to our question about additional terms to consider, the participants offered \formatTerm{heuristic}, \formatTerm{design pattern}, \formatTerm{requirement}, and \formatTerm{justification.} Through informal discussions, they also suggested \formatTerm{implication}. For the crowdsourced studies, discussed in the next section, we decided to incorporate all of these suggestions except \formatTerm{justification} because the term does not fit the guidelines space. A justification is not guidance about what should be done; it is an explanation of why something was done. We added the
definitions for the new terms at the bottom of \cref{tab:terms-final} using the same methodology as above. However, as the dictionaries did not offer definitions for \formatTerm{design pattern}, we based our initial definition on Dearden and Finlay's \cite{dearden2006PatternLanguagesHCI}.

\subsubsection{Limitations of This Initial Exploration}
This study provided useful initial structure, but it was limited
 by the small, non-representative sample of workshop attendees, and the dimensions were
 researcher-defined rather than data-driven. Thus, we conducted two follow-up studies: a larger-scale crowdsourced experiment to map the space of guidance terminology and understand its dimensions (\Cref{sec:experiments}), and a corpus analysis to understand actual term usage in the literature (\Cref{sec:termusage}).

\subsection{Crowdsourced Studies}%
\label{sec:experiments}

To test the extent to which
the space of guidance terminology from our exploratory study
represents the perspectives of broader visualization community members, we conducted a study surveying a larger population of visualization researchers to (1) map the space of terminology (mapping phase) in which participants made triadic similarity judgments and (2) understand the nature of the space's underlying dimensions by having participants rate terms on bipolar scales (rating phase). The experiment scripts, data, and analysis scripts for both phases of this study can be found in our GitHub repository (\href{https://github.com/SchlossVRL/vis_guidelines}{github.com/SchlossVRL/vis\_guidelines}).

\subsubsection{Participants}
To identify candidate participants, we first collected email addresses from IEEE VIS papers of the last five years using the \href{https://grobid.readthedocs.io/}{Grobid tool}. Next, we removed duplicates, incorrectly identified emails, and email lists. We considered excluding participants from Study~1, but opted to keep them in the pool as these studies consist of completely new tasks, and we did not want to limit input from members of the field. We then randomly divided the list into two halves: one that was invited to participate in the mapping phase, and the other that was subsequently invited to participate in the rating phase. We sent 387 invitations in the mapping phase and 395 in the rating phase (both numbers exclude bounced emails).
A total of 30 participants completed the mapping phase, and 42 participants completed the rating phase. In both phases, we excluded data from participants who reported having fewer than 1 year of experience in the visualization community or did not complete all trials (mapping phase: $n = 1$ excluded, rating phase: $n = 0$ excluded). For participants in the mapping phase, their reported gender included: 10 women, 18 men, 1 mostly male, and 1 who chose not to specify, and their reported years in the VIS community included: 9 with $>$ 1, 7 with $\geq$ 10, 9 with $\geq$ 10, 4 with $\geq$ 30, and 1 with 40+ years (min = 5, max = 40+). Reported gender identities in the rating phase included: 14 women, 26 men, 1 he/they, and one who chose not to specify, and reported years in the VIS community included: 15 with $>$ 1, 14 with $\geq$ 10, 10 with $\geq$ 10, and 3 with $\geq$ 30 (min = 3, max = 35). All participants gave informed consent, and the University of Utah IRB approved the experiment protocol.

\subsubsection{Mapping Phase}
\label{sec:study1}
This phase was analogous to the space exploration phase of Study 1 (\Cref{sec:space_explore}), but using a more rigorous embedding procedure.

\textbf{Procedure.} The stimulus set included 17 guidance-related terms from our initial list (\Cref{sec:term_generation}), expanded by terms suggested in survey feedback (\Cref{tab:terms-final}). The final set included: \formatTerm{design pattern}, \formatTerm{heuristic}, \formatTerm{requirement}, \formatTerm{implication}, \formatTerm{principle}, \formatTerm{rule of thumb}, \formatTerm{consideration}, \formatTerm{common practice}, \formatTerm{convention}, \formatTerm{recommendation}, \formatTerm{best practice}, \formatTerm{standard}, \formatTerm{norm}, \formatTerm{guardrail}, \formatTerm{paradigm}, \formatTerm{style guide}, and \formatTerm{guideline}.

We used a triadic similarity task to map the space of guidance terminology \cite{giallanza2024integrated, hebart2023things, jamieson2011low,  mukherjee2022color}.
While making their similarity judgments, participants were instructed to think back to times they had read the results or discussions of a VIS paper and came across words describing the contributions or results of the work.
On each trial, participants saw one target word and two choice words and selected the choice word that they felt was most similar to the target (see Supplemental \Cref{fig:tasks}A for an example trial).
They were informed there were no right or wrong answers for most trials and to go with their gut.
The only trials with `correct' answers were 10 catch trials in which one choice word was the same as the target. These catch trials were used to ensure participants were paying attention and understood their task, but were not included in analyses.

We collected 2,408 randomly sampled triplet trials for estimating our embeddings. Prior work established that at least $n\cdot d \cdot \log_2 n$ judgments are needed to estimate a stable $d$-dimensional embedding~\cite{jamieson2011low}, where $n$ is the number of items.
We collected 10 times this number to account for noise in judgments to arrive at a total of 2,408 judgments.
We distributed these trials across participants, each of whom completed 100 total trials---80 randomly sampled triplets used to fit the embedding, 10 validation triplets to estimate embedding fits, and 10 catch trials. Both crowdsourced experiments were presented through the browsers on participants' personal computers via jsPsych \cite{de2015jspsych, de2023jspsych}.

\textbf{Analysis and Results.}
We used the similarity judgments to estimate low-dimensional embeddings for each term such that terms frequently judged to be similar were embedded closer together in the space.
We estimated embeddings using the crowd kernel learning (CKL) algorithm \cite{tamuz2011adaptively}, which
does not require similarity judgments for all possible pairwise sets of terms, unlike other classic approaches like multidimensional scaling (MDS) \cite{torgerson1952multidimensional}.
Instead, by collecting a sufficiently large set of ordinal judgments (e.g., a duck is closer to a canary than to a horse), modern implementations of CKL \cite{sievert2023efficiently} can estimate a reliable embedding from a tractable number of human judgments.
We fit embeddings in spaces of up to five dimensions and measured each solution's performance with respect to dimensionality, thus identifying an appropriate dimensionality for the embedding (see Supplemental \Cref{sec:embeddings} for details). We observed that all solutions performed similarly in predicting the held-out human judgments.

\begin{figure*}[!t]
    \centering
    \includegraphics[width=.95\textwidth]{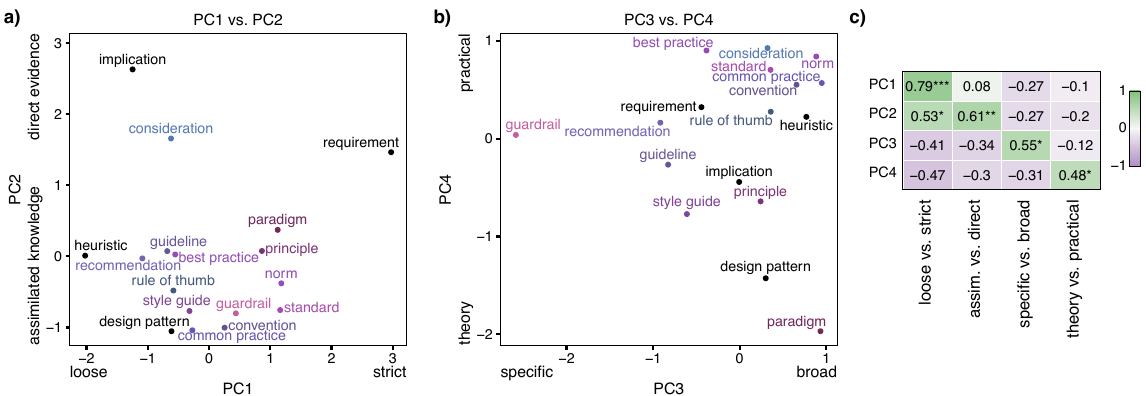}
    \caption{Guidance terms embedded in PC space for (a)~PC1 with PC2 and (b)~PC3 with PC4. The text color corresponds to \Cref{fig:term-uncertainty} from Study~1 (new terms for Study 2 are in black).  The axis labels of the PC plots are based on (c), which shows~Pearson's $r$ correlations between each rating scale and each PC: PC1 was strongly correlated with loose vs.\ strict, PC2 with direct evidence vs.\ assimilated knowledge, PC3 with broadness, and PC4 with theoretical vs.\ practical. $^*p<.05$, $^{**}p<.01$, $^{***}p<.001$.}
    \label{fig:pc-space-and-correlations}
    \vspace{-5mm}
\end{figure*}

To further assess the dimensionality of the guidance terminology space, we conducted principal component analysis (PCA) on the five-dimension solution. PCA revealed that the first four components accounted for 95\% of the variance in the embedding, with the fifth component contributing relatively little additional variance. These results suggested that the underlying structure of the guidance terminology space was likely 4D. We therefore retained a four-dimensional solution for subsequent analyses. To obtain an orthogonal set of components ordered by explained variance, we conducted PCA on the 4D embedding to interpret (\Cref{fig:pc-space-and-correlations}a,b; additional results can be found in the Supplemental Material).

The first principal component (PC) seemed to align with the Strictness dimension from the exploratory study in \Cref{sec:dagstuhl}. We also identified three potential dimensions that the additional components could represent: specific vs.\ broad (Coverage), theoretical vs.\ practical, and assimilated knowledge vs.\ direct evidence. To test how well these dimensions captured the variability in the guidance terminology space, we conducted the rating phase of the study.

\subsubsection{Rating Phase}
\label{sec:study2}
This phase was analogous to the dimension-rating task in Study 1 (\Cref{sec:study1-dimensions}), but with additional terms and scales.

\textbf{Procedure.}
Participants rated the 17 guidance terms along the following dimensions: loose vs.\ strict (Strictness in Study~1), specific vs.\ broad (Coverage in Study~1), theoretical vs.\ practical, and assimilated knowledge vs.\ direct evidence.

During the experiment, we instructed participants to think back to times when they had read the results or discussion of a VIS paper and encountered words describing the contributions or results of the work. Their task was to rate a list of such words, one at a time, on a series of scales. They were then shown the full list of terms to be rated and the full list of scales on which each term would be rated. They were asked to anchor the endpoints of each scale in the context of the terms \cite{palmer2013visual} by thinking about which terms they associate most strongly with each endpoint of each scale and to rate those words near the appropriate endpoints when encountered during the task. They were instructed to rate a word near the midpoint of a scale if they did not associate it strongly with either end of the scale. Participants would rate all of the words for a given scale, one at a time, before moving on to the next scale. Finally, before beginning, participants were told how to respond using the line-mark slider scale by sliding the cursor along the scale and clicking to record their response (see Supplemental \Cref{fig:tasks}B for an example trial).

We presented trials in a block-randomized design---all terms for a given scale were presented in random order within each scale, and the order of the scales was randomized. To reduce the risk of response-direction biases, we varied the endpoints of the scales so participants were equally likely to rate terms on reversed versions of scales. For analysis, we flipped the ratings on reversed scales to be consistent with original scales and scaled all ratings from 0 to 1.

\textbf{Analyses and Results.}
To test how well the scales described variability along the PCs from the mapping phase, we correlated the mean ratings for each term along each scale with the coordinates for each term along each PC. \Cref{fig:pc-space-and-correlations}c shows the resulting correlation matrix. PC1 was strongly correlated with loose vs.\ strict, PC2 with direct evidence vs.\ assimilated knowledge, PC3 with specific vs.\ broad, and PC4 with theoretical vs.\ practical. Thus, similar to Study~1, strictness was the primary dimension of the term space (PC1), and coverage was also a relevant dimension (PC3), but this study also revealed two additional dimensions.

 \Cref{fig:teaser} (right) shows the mean ratings for each term along each of the four scales. For example, for loose vs.\ strict, looser terms included \formatTerm{rule of thumb} and \formatTerm{common practice}, whereas stricter terms included \formatTerm{standard} and \formatTerm{requirement}; for assimilated knowledge vs.\ direct evidence, assimilated knowledge terms included \formatTerm{rule of thumb} and \formatTerm{convention}, whereas direct evidence terms included \formatTerm{requirement} and \formatTerm{implication}; for specific vs.\ broad, specific terms included \formatTerm{requirement} and \formatTerm{recommendation}, whereas broad terms included \formatTerm{common practice} and \formatTerm{paradigm}; and for theory vs.\ practical, theory terms included \formatTerm{principle} and \formatTerm{paradigm}, whereas practical terms included \formatTerm{common practice} and \formatTerm{best practice}.

 These findings can be used to help researchers, educators, and designers select terminology that best aligns with their intended message when communicating about visualization guidance.

\begin{figure*}[!t]
  \centering
  \begin{subfigure}[b]{0.36\textwidth}
    \centering
    \includegraphics[width=\textwidth]{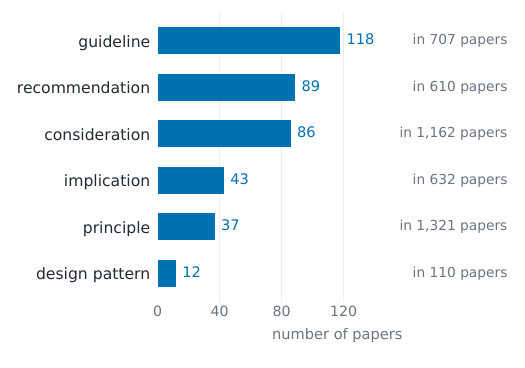}
    \caption{Term distribution}
    \label{fig:term-classification}
  \end{subfigure}
  \hfill
  \begin{subfigure}[b]{0.32\textwidth}
    \centering
    \includegraphics[width=\textwidth]{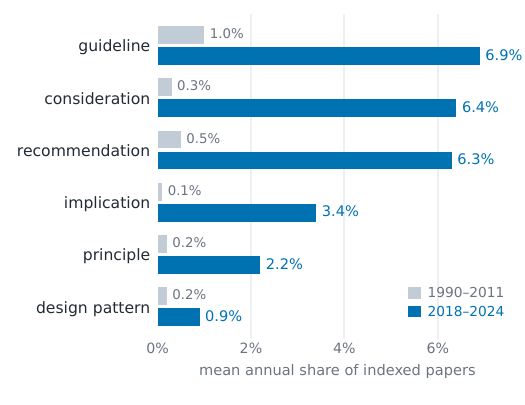}
    \caption{Temporal trends}
    \label{fig:term-trends}
  \end{subfigure}
  \hfill
  \begin{subfigure}[b]{0.3\textwidth}
    \centering
    \includegraphics[width=\textwidth]{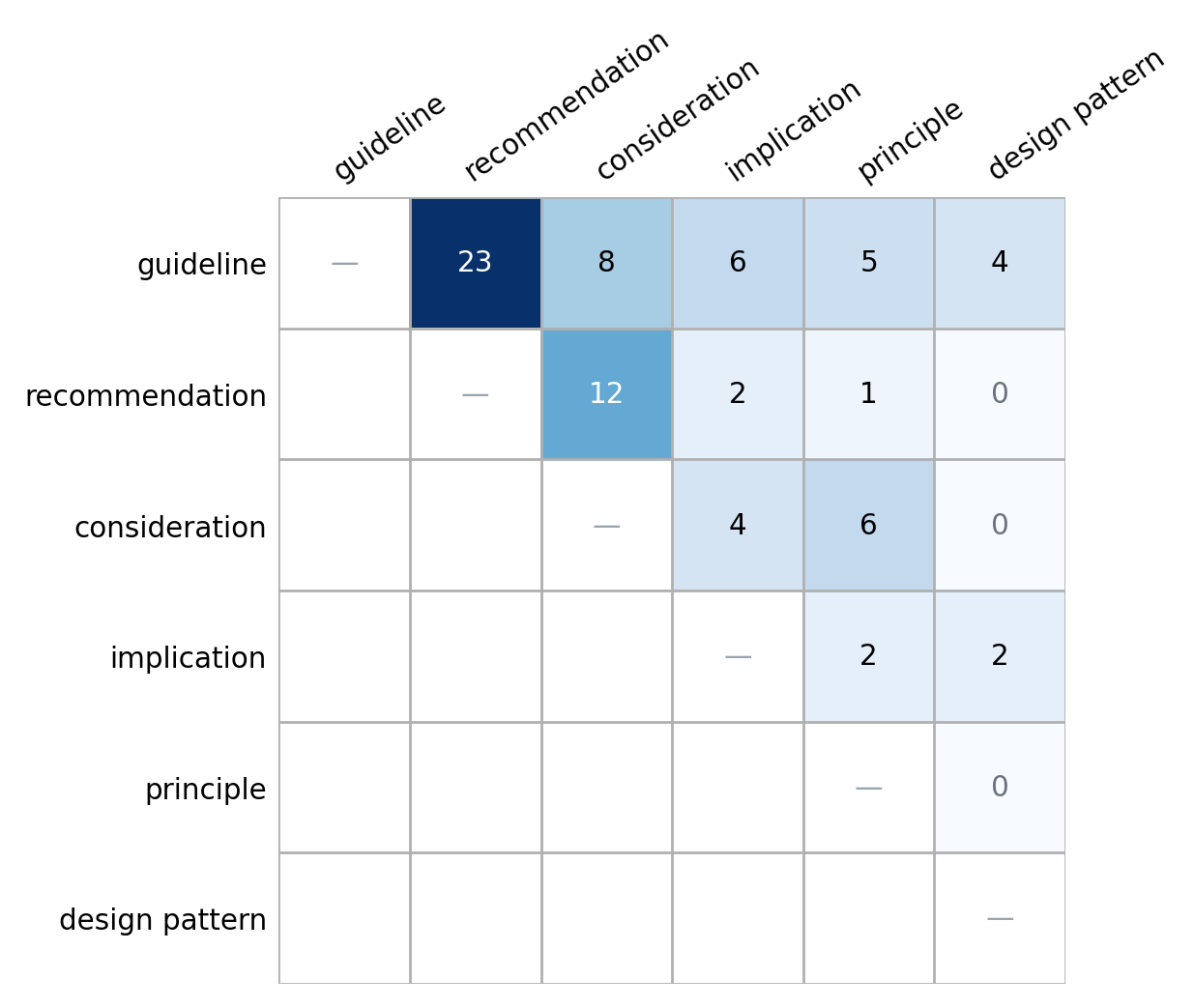}
    \caption{Term co-occurrence}
    \label{fig:term-cooccurrence}
  \end{subfigure}
  \caption{Guidance-adjacent term usage in the IEEE VIS corpus: (a)~Guidance-adjacent use of the six most frequent terms,
  (b)~their comparison for the 1990--2011 and 2018--2024 periods, and
  (c)~pairwise co-occurrence of guidance-adjacent terms within papers.
  }
  \label{fig:term-usage-combined}
  \vspace{-4mm}
\end{figure*}

\section{Term Usage in Visualization Literature}
\label{sec:termusage}
We now move from perceived meaning to observed use, examining how  terms appear in published VIS research. We focus on guidance attributed to a paper's own findings, analysis, or design process, and set aside prior work, technical meanings, and unrelated uses. We characterize this usage rather than judge whether an author selected the most appropriate term. The results can be explored at
 \href{https://vis-guidelines.netlify.app/}{vis-guidelines.netlify.app}.

\subsection{Corpus and Method}
We analyzed the full text of 3,877 IEEE VIS papers published from 1990 to 2024, collected from
 \href{https://vispubdata.org}{vispubdata.org}%
~\cite{isenberg2016vispubdata}. We retrieved the exact singular and plural forms of the 18 terms in our terminology analysis and retained the surrounding passage, paper metadata, and document location for each match. This process yielded 32,629 occurrences across 10,838 term--paper pairs.
We classified each occurrence as \emph{guidance-adjacent}, \emph{ambiguous}, or \emph{likely non-guidance}. Guidance-adjacent cases described design or practice guidance attributed to a paper's work and presented as relevant beyond the immediate system or study. Ambiguous cases lacked enough context to determine the term's function, source, or intended scope. Likely non-guidance cases included technical or idiomatic meanings, bibliographic occurrences, guidance attributed to prior work, system-generated recommendations, and other unrelated uses.

We used Claude Opus 4.7 to assign an initial label to each occurrence according to these criteria. We then manually reviewed every occurrence initially labeled guidance-adjacent and corrected misclassifications. At the term--paper level, we assigned the strongest applicable label: guidance-adjacent if the pair contained at least one verified guidance-adjacent occurrence, ambiguous if it contained no guidance-adjacent occurrence but at least one ambiguous occurrence, and likely non-guidance otherwise. After classification and manual verification, we identified 625 guidance-adjacent occurrences across 319 of the 3,877 papers.

The supplemental website (\href{https://vis-guidelines.netlify.app/}{vis-guidelines.netlify.app}) provides the classification criteria, term-specific notes, representative and borderline examples, aggregation rule, and final occurrence-level classifications used in our analysis.

\subsection{Results}
\label{sec:term-usage-findings}

\subsubsection{Distribution of Guidance-Adjacent Terms}

The corpus of use was much narrower than our terminology landscape in \Cref{sec:empirical}. Authors used a small set of terms to describe guidance in their own work, even though those terms may differ in how visualization researchers understand them. Overall word frequency also differed from guidance-adjacent use, and most papers relied on one guidance-related label rather than using several labels.

\textbf{Concentration in a small set of terms.}
We verified 625 guidance-adjacent occurrences in 319 papers, or 8.2\% of the corpus. These occurrences formed 392 term--paper pairs, of which 385, or 98.2\%, involved only six terms: \formatTerm{guideline}, \formatTerm{recommendation}, \formatTerm{consideration}, \formatTerm{implication}, \formatTerm{principle}, and \formatTerm{design pattern} (\Cref{fig:term-classification}). \formatTerm{Guideline} had the largest number of guidance-adjacent papers at 118, followed by \formatTerm{recommendation} at 89, and \formatTerm{consideration} at 86. \formatTerm{Best practice}, \formatTerm{heuristic}, and \formatTerm{rule of thumb} together accounted for only seven additional pairs, while the remaining nine terms had no verified guidance-adjacent use.
This concentration does not mean that the six terms describe the same kind of outcome. \Cref{sec:empirical} shows that they differ in perceived strictness, breadth, evidentiary basis, and practical orientation. A \formatTerm{consideration} derived from a design process, an \formatTerm{implication} drawn from an empirical result, and a \formatTerm{recommendation} directed toward practitioners can all appear as outcomes of a paper, but they communicate different expectations about how broadly, directly, or strongly the result should guide subsequent work. The corpus therefore shows that a small vocabulary is used to communicate several distinct forms of guidance-related knowledge.

\textbf{Prevalence and guidance-adjacent use.}
A term's overall prevalence did not indicate how often authors used it to present guidance from their own work. \formatTerm{Standard} appeared in 2,176 papers (56.1\% of the corpus) and \formatTerm{requirement} in 1,693 papers (43.7\%), yet we retained no guidance-adjacent occurrence for either term. In the reviewed passages, \formatTerm{requirement} usually referred to specifications for the system or study at hand rather than an outcome intended to inform work beyond that immediate context.
By comparison, guidance-adjacent uses appeared in 16.7\% of papers for \formatTerm{guideline}, in 14.6\% for \formatTerm{recommendation}, and in 10.9\% for \formatTerm{design pattern}. The corresponding shares were lower for \formatTerm{consideration}, \formatTerm{implication}, and \formatTerm{principle}, even though some of these words appeared more often overall.

These differences separate the presence of a word from the role that it serves in a paper. Terms such as \formatTerm{standard} and \formatTerm{requirement} were common because they frequently described technical criteria, study conditions, or system specifications. Terms such as \formatTerm{guideline} and \formatTerm{recommendation} were less common overall but more often used to present guidance-related outcomes. Whether a term signals a guidance contribution depends on the surrounding claim, not simply on how often the word appears in the literature.

\subsubsection{Temporal Patterns}

Each of the six most frequent terms had a higher mean annual share of guidance-adjacent use in 2018--2024 than in 1990--2011 (\Cref{fig:term-trends}): from 1.0\% to 6.9\% for \formatTerm{guideline}, from 0.5\% to 6.3\% for \formatTerm{recommendation}, and from 0.3\% to 6.4\% for \formatTerm{consideration}. \formatTerm{Implication}, \formatTerm{principle}, and \formatTerm{design pattern} also had higher shares in the later period, although they remained less common.

These period averages do not establish a steady year-over-year increase, and the corpus does not explain why guidance-adjacent language became more common. The pattern is nevertheless broader than the increasing use of \formatTerm{guideline} alone. Several terms became more visible as labels for outcomes of a paper. The later literature continued to use multiple labels rather than converging on a single term for guidance-related contributions.

\subsubsection{Within-Paper Patterns}

Most papers used only one guidance-adjacent term. Of the 319 papers, 252 (79.0\%) used one term, 61 used two, and six used three. The most frequent combination was \formatTerm{guideline} and \formatTerm{recommendation}, found in 23 papers, followed by \formatTerm{consideration} and \formatTerm{recommendation}, found in 12 (\Cref{fig:term-cooccurrence}).
The presence of two terms in a single paper does not imply that the authors treated them as synonyms, as the terms may describe different findings, inputs, or outputs. Still, co-use provides some evidence that authors alternated among several labels when discussing guidance-related outcomes. The corpus, therefore, does not point to indiscriminate switching as the main issue. Instead, a single label often serves as the primary cue to the status of a claim, even though \Cref{sec:empirical} shows that readers may associate that label with different levels of strictness, breadth, evidence, and practical relevance.

\subsection{Implications and Limitations}
\label{sec:term-usage-implications}

\subsubsection{Term Use and Interpretation}

Our corpus analysis shows that authors used only a narrow subset of terms when describing guidance from their own work. Six terms accounted for 98.2\% of the guidance-adjacent term--paper pairs, and 79.0\% of the papers used only one such term.
For example, the varied uses of \formatTerm{guideline} illustrate the limits of relying on the label alone. Marschner and Lobb described quantitative metrics for evaluating reconstruction filters as providing ``a useful guideline''~\cite{marschner1994evaluation}, whereas Bederson and Boltman used the term for advice derived from an experiment, referring to ``preliminary design guidelines''~\cite{bederson1999animation}. Djurcilov and Pang applied ``a general guideline'' to a conditional recommendation about when to use scattered-data methods~\cite{djurcilov1999visualizing}. Stasko and Zhang presented ``design guidelines'' developed from system experience, experimental observations, and an analysis of related systems~\cite{stasko2000focuscontext}. In these papers, \formatTerm{guideline} refers to evaluative criteria, tentative empirical advice, a conditional prescription, and guidance developed from several forms of evidence.

These examples do not establish that the term was used incorrectly. They show that \formatTerm{guideline} alone does not specify the form of the outcome, the evidence supporting it, its intended scope, or how strictly it should be followed. The same concern applies to \formatTerm{consideration}, \formatTerm{implication}, \formatTerm{recommendation}, \formatTerm{principle}, and \formatTerm{design pattern}. Such terms can help characterize a contribution, but they do not replace an account of what the claim applies to, where it comes from, whom it is intended to inform, and how it should guide action. Our analysis therefore concerns the communication of guidance rather than the correctness of individual word choices. It also cannot explain why particular terms were favored; familiarity, disciplinary convention, reviewer expectations, and fit to the contribution remain possible explanations.

\subsubsection{Sources of Guidance}

We retained guidance presented as an outcome of a paper's findings, analysis, or design process. We checked how the term was attributed, not whether it rested on a single form of evidence. Some cases drew on an experiment, while others combined empirical observations, design experience, and analysis conducted within the paper. We excluded passages that only attributed guidance to prior work, professional practice, community consensus, or another external source. The analysis, therefore, primarily captures the direct-evidence side of the distinction in \Cref{sec:empirical}, rather than the broader process by which evidence and experience accumulate into established guidance.

\section{Recommendations}
\label{sec:recommendations}

Our results highlighted a divergence in how visualization authors understand guidance-related terms. This divergence is problematic for research progress and the practical impact of visualization work. When researchers and authors have diverging understandings of what authors contributed and what they expected in terms of a contribution, frustration on either side is almost inevitable in the publication process. When students search for clear rules to follow but encounter only loose suggestions, they may feel lost. Furthermore, if guidance is understood primarily as a means of informing practice, researchers have argued that we first need to determine what type of guidance practitioners actually need \cite{parsons:2022:ConsideringRoleGuidelines}.

It is clear that as we communicate about ways to approach studying and producing effective visualizations, we have to be mindful of the implications of the terms we use. Below, we outline four categories of \formatTerm{recommendations} which should be seen as ``practical, experience-based, rather loosely binding yet specific suggestions'' according to our study results:

\textbf{Authors.} Authors of visualization papers should think carefully about what they discovered and choose terms that really fit the contribution. For example, a paper describing a new design space might not be able to offer strict requirements or even loose considerations without direct evidence, but could easily describe common practice or make recommendations for new avenues to explore. \Cref{fig:teaser} and \Cref{tab:terms-final} can help with choosing the right terminology, or authors could use them as a basis for adapting and describing a changed use of the term. In addition, our axes can serve as considerations for describing the scope of contributions: are they meant to be rather loose suggestions or strict requirements to follow, can they be broadly applicable or only to the specific context of the paper, or are they meant to advance theory or practice?

\textbf{Reviewers.} If reviewers are aware of the nuances of guidance concepts in visualization research and hold a shared understanding of terminology, that will help ensure their feedback aligns with the paper's content and goals. Moreover, this understanding will help reviewers avoid asking for inappropriate revisions, such as asking for stricter terms, such as \formatTerm{guidelines}, even though only loose terms would make sense given the research presented. It may be compelling for reviewers to push authors to stricter, broader, and more actionable conclusions of their work. If the research outcomes of a paper under review provide evidence to support such conclusions, then, of course, they should be incorporated in the revision process. However, reviewers are advised not to push too far, as doing so could result in published papers with potentially misleading or unreliable messages.

\textbf{Research community.}
Reliable descriptions of research outcomes are key to progress in the research community as we continue building on the existing body of published work. Establishing a clear, shared terminology may improve our ability to assemble, merge, and synthesize knowledge from collections of papers (survey papers, textbooks, course notes, etc.). This effort can pave a way toward theory, following a path from single studies to meta-analyses to principles---at least for some facets of visualization. Some pathways toward visualization theory are discussed by Chen et al.~\cite{chen:2017:PathwaysTheoreticalAdvances}, see also ~\cite{schloss2025perceptual}.
The findings of our paper can serve as a basis for the research community to adopt a shared, nuanced terminology, but ongoing discussion will be needed until a unified understanding is established. This process might include an iterative approach of evolving and reviewing publication expectations, and could be fostered by future workshops, courses at conferences, or other community events.

\textbf{Educators.} Teachers and other educators of visualization research or practice are encouraged to clearly articulate what kind of messages they want to communicate, e.g., to students or outside practitioners. In education, there may be less shared terminology and, more importantly, less shared background knowledge than in communication among researchers within the visualization research community. Therefore,  even more effort may be needed to clarify the meaning of terminology, e.g., by explaining or defining terms. With an appropriate understanding, it will be easier to manage the expectations of students or practitioners, thus fostering the actionability of visualization research findings.

\section{Conclusion}
\label{sec:conclusion}

The main goal of this paper is to raise awareness of the terminology used in the visualization guidance landscape and to shed some light on how terms are interpreted and used in the visualization research community. We based our discussion on a diverse and complementary set of sources: an exploratory study with experts at a Dagstuhl Seminar, a crowdsourcing study (with separate mapping and rating phases) that targeted input from the broader visualization community, and an investigation of the use of terminology in published VIS research. Our results are summarized in the ratings shown in \cref{fig:teaser} and our list of terms and their definitions in \cref{tab:terms-final}.

We decided to apply our findings to our own paper: After discussion within the group of authors, we converged to naming \cref{sec:recommendations} \formatTerm{Recommendations}, based on our judgment regarding strictness (here being slightly toward the loose end) and being rather specific, practical, and mostly based on assimilated knowledge (observations from the literature and feedback from many experts in the field). However, we want to point out that there could have been other valid choices for the term used in our case. Also, we acknowledge that our list of terms is not comprehensive, e.g., additional useful terms could be \formatTerm{rules}, \formatTerm{constraints}, or \formatTerm{suggestions}.

Therefore, in general, we advocate for some flexibility, to be able to accommodate the many different facets and nuances in guidance results of visualization research, echoing the call for supporting diversity in visualization contributions by Lee et al.~\cite{lee2019broadening}. Through a negotiation
process within the community, a better, shared understanding of terminology can be achieved, thus helping how we communicate guidance-related findings.

\section*{AI Use Statement}
Claude 4.8 Opus helped generate the illustrative artwork in \cref{fig:teaser} (the data plots were not AI-generated), assign labels (i.e., guidance-adjacent) to occurrences of terms in our corpus analysis in \cref{sec:termusage}, and provided editorial and structural support in preparing the manuscript (this assistance did not generate the research content, findings, or conclusions). GPT 5.3 and Claude 4.8 Opus helped debug and write experiment code and debug figure style code. ChatGPT helped with definitions for guidance-related terminology in \cref{sec:dagstuhl}. All AI generated content was reviewed by the authors for accuracy.

\acknowledgments{
This work stemmed from Dagstuhl Seminar 25232, ``Navigating the Maze of Guidelines to Unify Visualization Design Recommendations.''
We thank Cindy Bearfield Xiong for participating in early stages of this work, and thank all seminar participants and crowdsource study participants. KM was supported by a seed grant from the Stanford Accelerator for Learning and National Science Foundation (NSF) award \#2400471, KBS and ALC by NSF award \#2419493, and DW by the Deutsche Forschungsgemeinschaft (DFG, German Research Foundation)--Project-ID 251654672--TRR 161.

}

\bibliographystyle{abbrv-doi-hyperref-narrow}

\bibliography{main}

\clearpage
\appendix
\renewcommand*{\thesection}{S}
\setcounter{figure}{0}
\renewcommand{\thefigure}{S.\arabic{figure}}
\setcounter{table}{0}
\renewcommand{\thetable}{S.\arabic{table}}
\section{Supplementary Material}\label{sec:supplementary}

\subsection{Terms and Definitions from the Initial Exploratory Study}

\begin{figure}[!h]
    \centering
    \includegraphics[width=.7\linewidth]{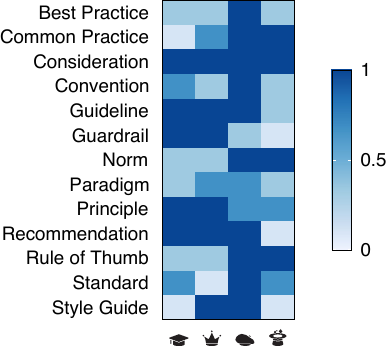}
    \caption{Frequency of responses for the likely sources of each term from \Cref{sec:study1-dimensions}. The colors correspond to the icons encodings in \Cref{tab:terms-final} and \Cref{tab:terms} (\includegraphics[height=.9em]{phd-hat.png}~empirical evidence;
    \includegraphics[height=1em]{crown.png}~eminent leader;
    \includegraphics[height=.8em]{artist-hat.png}~art \& design community;
    \includegraphics[height=1em]{magician-hat.png}~emergent from unknown source).
     }
    \label{fig:origin_mat}
    \vspace{-4mm}
\end{figure}

\Cref{tab:terms} provides the terms and their preliminary definitions, according to the exploratory study at the Dagstuhl Seminar (see \cref{sec:dagstuhl}). The table also contains the surveyed sources for each of the terms. \Cref{fig:origin_mat} shows the frequency of responses for the likely sources of each term described in \cref{sec:study1-dimensions} and encoded in the icons in \cref{tab:terms-final} and \cref{tab:terms}.

\begin{table}[h!]
    \caption{A list of terms and definitions from our initial understanding of the landscape
    of guidance-related terminology.
    Icons indicate likely sources of each term, determined using survey data from \cref{sec:dagstuhl}, Exploration of Guidance Terminology:
    \includegraphics[height=.9em]{phd-hat.png}~empirical evidence;
    \includegraphics[height=1em]{crown.png}~eminent leader;
    \includegraphics[height=.8em]{artist-hat.png}~art \& design community;
    \includegraphics[height=1em]{magician-hat.png}~emergent from unknown source.
    Icon darkness reflects prevalence in survey responses from most to least prevalent (\opbulletone~\opbullettwo~\opbulletthree~\opbulletfour).}
    \label{tab:terms}
    \centering
    \small
    \begin{tabularx}{\columnwidth}{@{}>{\raggedleft\arraybackslash}p{7.5em}>{\raggedright\arraybackslash\hyphenpenalty=10000\exhyphenpenalty=10000}X@{\hspace{3pt}}c@{\hspace{3pt}}c@{\hspace{3pt}}c@{\hspace{3pt}}c@{}}
       \textbf{Principle}
         & A fundamental primary or general law or truth from which \textit{others} are derived
         & {\DBIconemp} & {\DBIconlead} & {\LBIconart} & {\LBIconunk}\\[2pt]
       \textbf{Rule of Thumb}
         & A rough method of procedure based on experience and common sense
         & {\BIconemp} & {\BIconlead} & {\DBIconart} & {\DBIconunk}\\[2pt]
       \textbf{Consideration}
         & Something to be kept in mind or to be given careful thought
         & {\DBIconemp} & {\DBIconlead} & {\DBIconart} & {\DBIconunk}\\[2pt]
       \textbf{Common Practice}
         & A typical way of doing something
         & {\WIconemp} & {\LBIconlead} & {\DBIconart} & {\DBIconunk}\\[2pt]
       \textbf{Convention}
         & An established and commonly accepted way of doing something
         & {\LBIconemp} & {\BIconlead} & {\DBIconart} & {\BIconunk}\\[2pt]
       \textbf{Recommendation}
         & A suggestion or piece of advice about what someone should do, choose, or consider
         & {\DBIconemp} & {\DBIconlead} & {\DBIconart} & {\WIconunk}\\[2pt]
       \textbf{Best Practice}
         & A widely accepted method or technique for reliably producing desirable results
         & {\BIconemp} & {\BIconlead} & {\DBIconart} & {\BIconunk}\\[2pt]
       \textbf{Standard}
         & An established set of criteria intended to ensure desirable results
         & {\LBIconemp} & {\WIconlead} & {\DBIconart} & {\LBIconunk}\\[2pt]
       \textbf{Norm}
         & A commonly accepted convention or practice
         & {\BIconemp} & {\BIconlead} & {\DBIconart} & {\DBIconunk}\\[2pt]
       \textbf{Guardrail}
         & A constraint that helps prevent undesirable results
         & {\DBIconemp} & {\DBIconlead} & {\BIconart} & {\WIconunk}\\[2pt]
       \textbf{Paradigm}
         & A prevailing framework or model that shapes how problems are approached and solutions are developed
         & {\BIconemp} & {\LBIconlead} & {\LBIconart} & {\BIconunk}\\[2pt]
       \textbf{Style Guide}
         & A set of rules or guidelines that help maintain consistency in the way something is presented
         & {\WIconemp} & {\DBIconlead} & {\DBIconart} & {\WIconunk}\\[2pt]
       \textbf{Guideline}
         & Suggestion or recommendation to help steer actions or decisions without being a strict rule
         & {\DBIconemp} & {\DBIconlead} & {\DBIconart} & {\BIconunk}\\

    \end{tabularx}
\end{table}

\subsection{Example trials the Mapping and Rating Phases of the Crowdsourced Study}
\label{sec:crowdsourse-tasks}

\begin{figure}[!h]
    \centering
    \includegraphics[width=.9\linewidth]{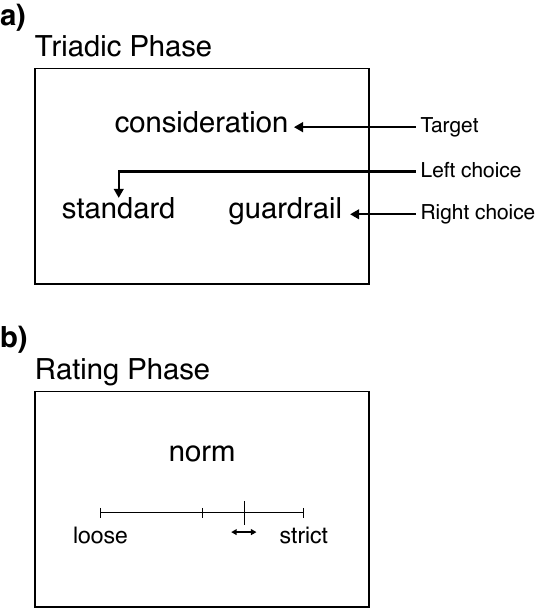}
    \caption{Example (A) triadic similarity judgment trial and (B)~rating trial for the crowdsourced study (\cref{sec:study1} and \cref{sec:study2}).
     }
    \label{fig:tasks}
    \vspace{-4mm}
\end{figure}

\Cref{fig:tasks} depicts example trials for both the triadic similarity task of the mapping phase and the rating task for the rating phase of the crowdsourced study. Descriptions of task instructions can be found in \cref{sec:study1} (triadic similarity judgment task) and \cref{sec:study2} (rating task) of the main text.

\subsection{Embeddings from the Mapping Phase of the Crowdsourced Study}
\label{sec:embeddings}

Here, we provide more information regarding the the mapping phase of the crowdsourced study (see \cref{sec:experiments} of the main paper).

\begin{figure}[!h]
    \vspace{2pt}
    \centering
    \includegraphics[width=.9\linewidth]{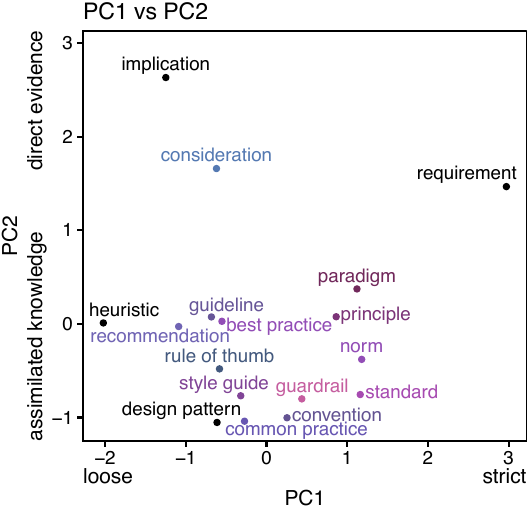}
    \caption{PC1 (loose vs.\ strict) vs.\ PC2 (assimilated knowledge vs.\ direct evidence) as depicted in \cref{fig:pc-space-and-correlations}.
     }
    \label{fig:PC1_PC2}
\end{figure}

\begin{figure}[!h]
    \centering
    \includegraphics[width=.9\linewidth]{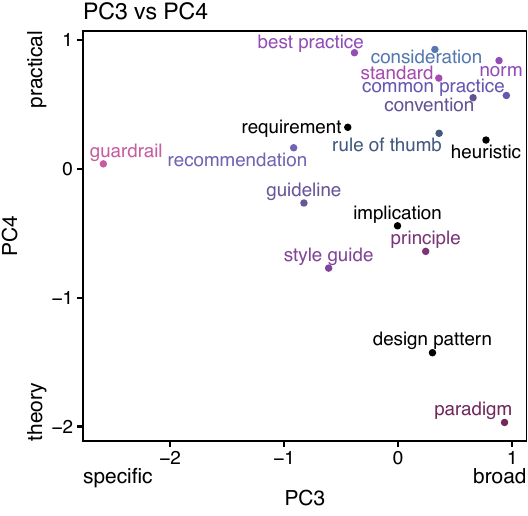}
    \caption{PC3 (specific vs.\ broad) vs.\ PC4 (theory vs.\ practical) as depicted in \cref{fig:pc-space-and-correlations}.
     }
    \label{fig:PC3_PC4}
    \vspace{-4mm}
\end{figure}

\begin{figure*}[!h]

\begin{minipage}[t]{0.475\textwidth}
  \centering
  \includegraphics[width=\linewidth]{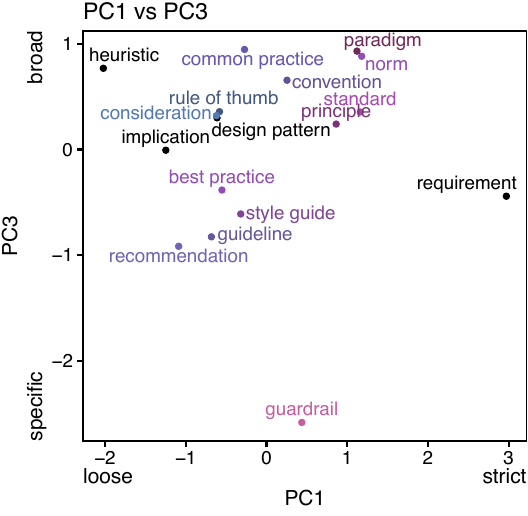}
  \caption{PC1 (loose vs.\ strict) vs.\ PC3 (specific vs.\ broad).}
  \label{fig:PC1_PC3}
\end{minipage}
\hfill
\begin{minipage}[t]{0.475\textwidth}
  \centering
  \includegraphics[width=\linewidth]{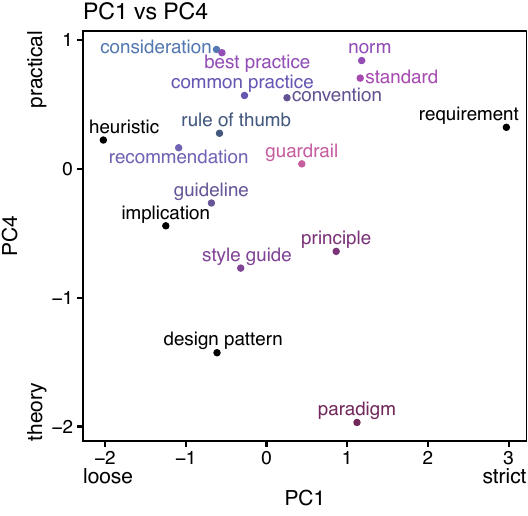}
  \caption{PC1 (loose vs.\ strict) vs.\ PC4 (theory vs.\ practical).}
  \label{fig:PC1_PC4}
\end{minipage}

\vspace{2em}
\begin{minipage}[t]{0.475\textwidth}
  \centering
  \includegraphics[width=\columnwidth]{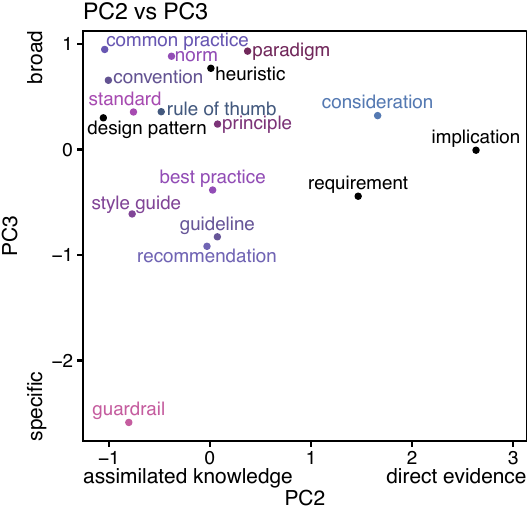}
  \caption{PC2 (assimilated knowledge vs.\ direct evidence) vs.\ PC3 (specific vs.\ broad).}
  \label{fig:PC2_PC3}
\end{minipage}
\hfill
\begin{minipage}[t]{0.475\textwidth}
  \centering
  \includegraphics[width=\columnwidth]{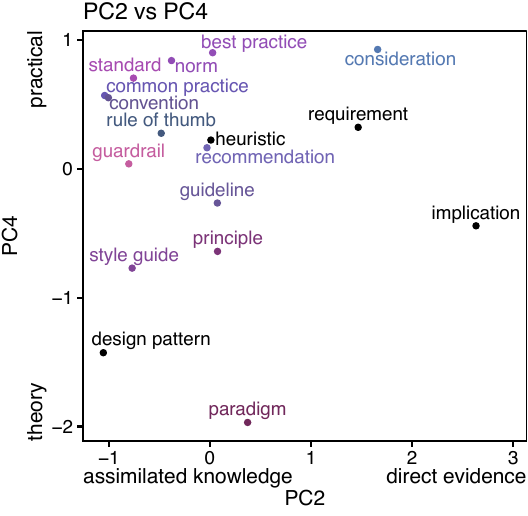}
  \caption{PC2 (assimilated knowledge vs.\ direct evidence) vs.\ PC4 (theory vs.\ practical).}
  \label{fig:PC2_PC4}
\end{minipage}

\end{figure*}

We include the guidance terms plotted as seen in \cref{fig:pc-space-and-correlations} (see \cref{fig:PC1_PC2} and \cref{fig:PC3_PC4}), and provide \Crefrange{fig:PC1_PC3}{fig:PC2_PC4}, where we plot guidance terms in each remaining pairwise combination of the four principal components (PC1: loose vs.\ strict, PC2: assimilated knowledge vs.\ direct evidence, PC3: specific vs.\ broad, PC4: theory vs.\ practical) alongside the variance explained and model loss used to select the embedding dimensionality.

\begin{figure}[!h]
    \centering
    \includegraphics[width=\linewidth]{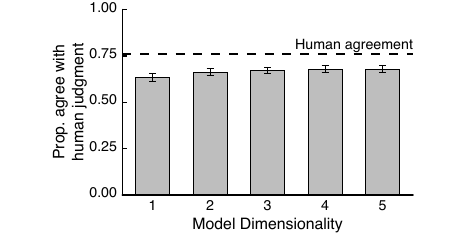}
    \caption{Comparison of model agreement with the majority of human judgments across embedding dimensionalities. The dashed line indicates the proportion of human judgments that agreed with the majority response for a set of held-out validation trials consistent across participants.
     }
    \label{fig:model_agree}
    \vspace{-4mm}
\end{figure}

For the underlying analysis, we fit embeddings at five dimensionalities (one, two, three, four, and five where five was the highest dimensionality for which we could estimate a stable embedding) and computed how well each solution predicted held out similarity judgments.
To measure each solution's performance, we first calculated the majority response for each triplet combination
to serve as the human response. We then calculated the Euclidean distance for each embedding space between the target word and its two options for each trial. We defined the option with the smaller Euclidean distance from the target as the model's response for which option was ``more similar.'' We then calculated the proportion of responses which aligned with the human majority response for each embedding solution. To establish a baseline for comparison with the variability of human responses, we calculated the proportion agreement across participants with the majority response for a set of 10 held-out validation trials judged by all participants (see \cref{fig:model_agree}).

\begin{figure}[!h]
    \centering
    \includegraphics[width=\linewidth]{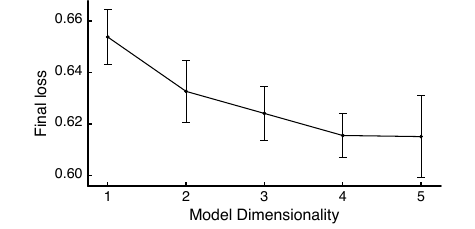}
    \caption{Final model loss across embedding dimensionalities. Lower final loss indicates better model fit to the data. We estimated 10 embeddings for each dimensionality with stochastic sampling and computed 95\% confidence intervals for the resulting average final loss scores.
     }
    \label{fig:loss}
    \vspace{-4mm}
\end{figure}

\begin{figure}[!h]
    \centering
    \includegraphics[width=\linewidth]{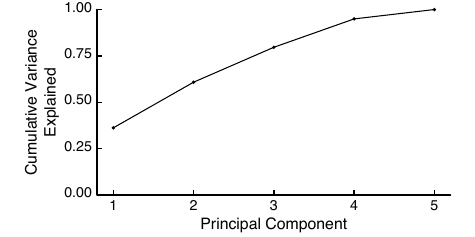}
    \caption{Cumulative variance explained by each principal component when conducting PCA on a 5-dimension embedding solution.
     }
    \label{fig:reduction}
\end{figure}

We observed that all solutions performed similarly in predicting the held-out human judgments. When comparing the final loss of each model, we found that the 4-dimensional model had the best fit to the data (lowest final loss), and the addition of a 5th dimension did not improve performance (see \cref{fig:loss}). To further assess the dimensionality of the guidance terminology space, we conducted principal component analysis (PCA) on the 5-dimension solution (see \cref{fig:reduction}). The PCA revealed the first four components accounted for 95\% of the variance in the embedding, with the fifth component contributing relatively little additional variance. These results suggested that the underlying structure of the guidance terminology space was likely four-dimensional.

\clearpage

\end{document}